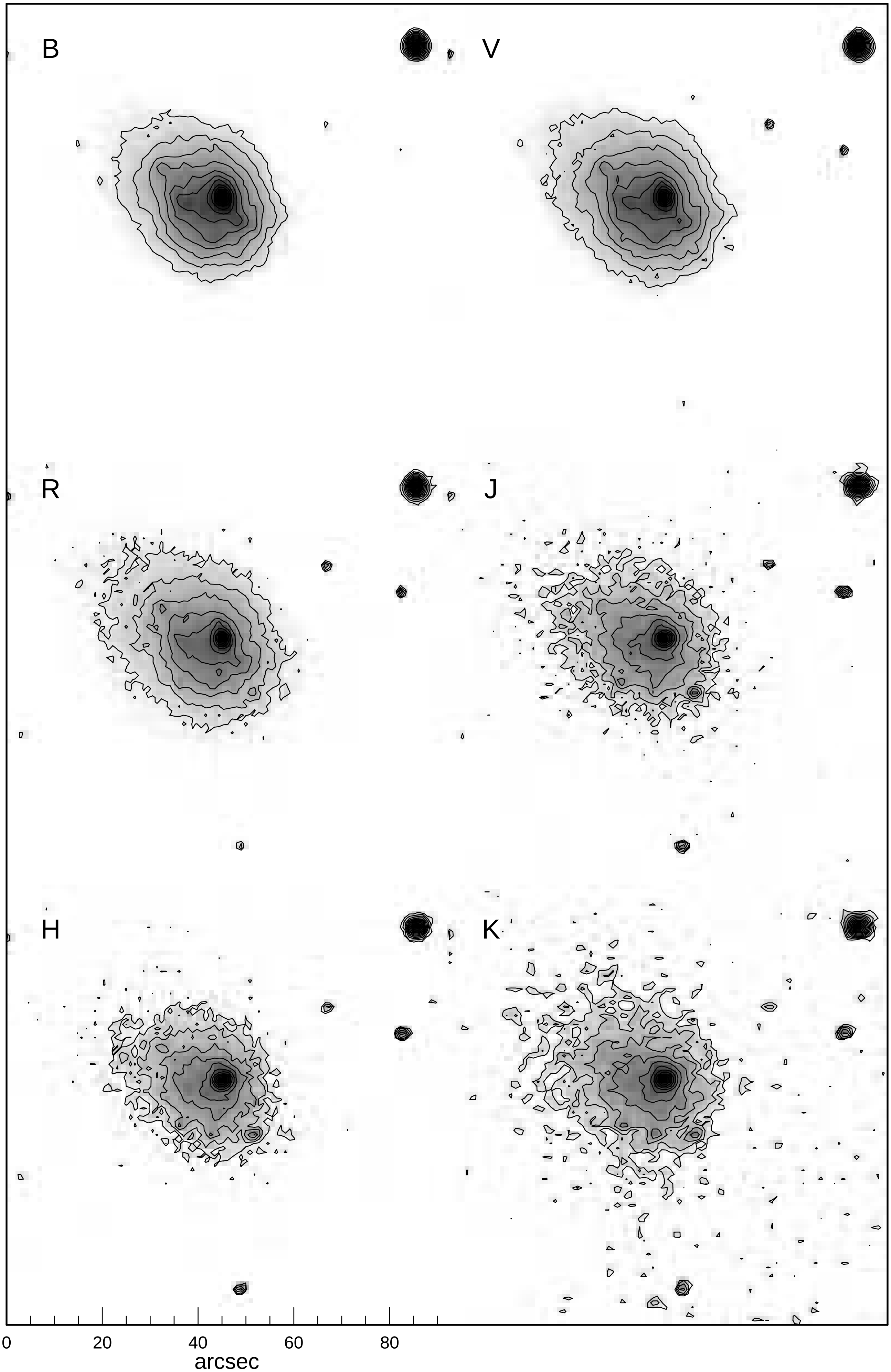

B
V
R
J
H
K
0
20
40
60
80
arcsec

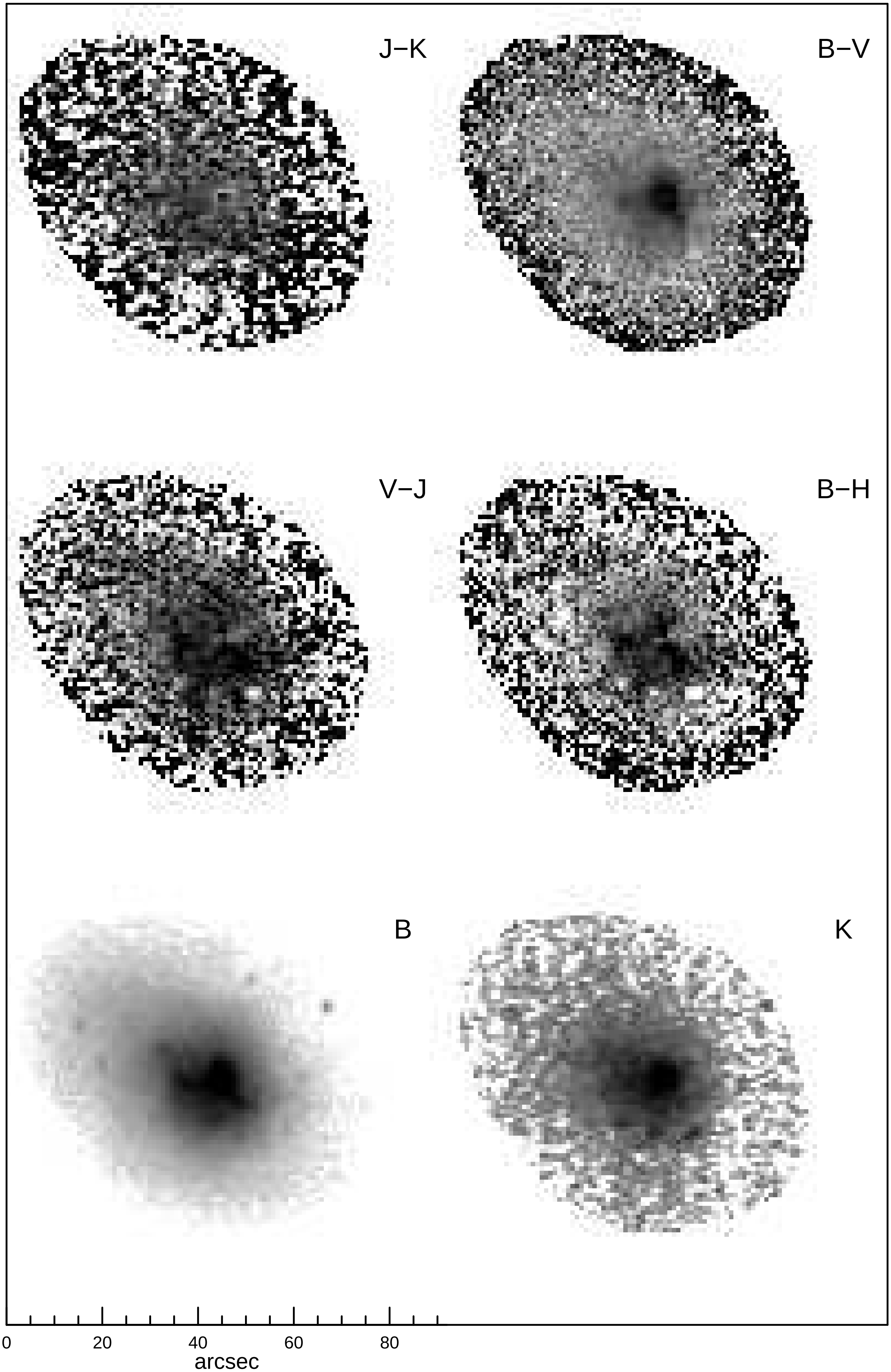

J−K
B−V
V−J
B−H
B
K
0
20
40
60
80
arcsec

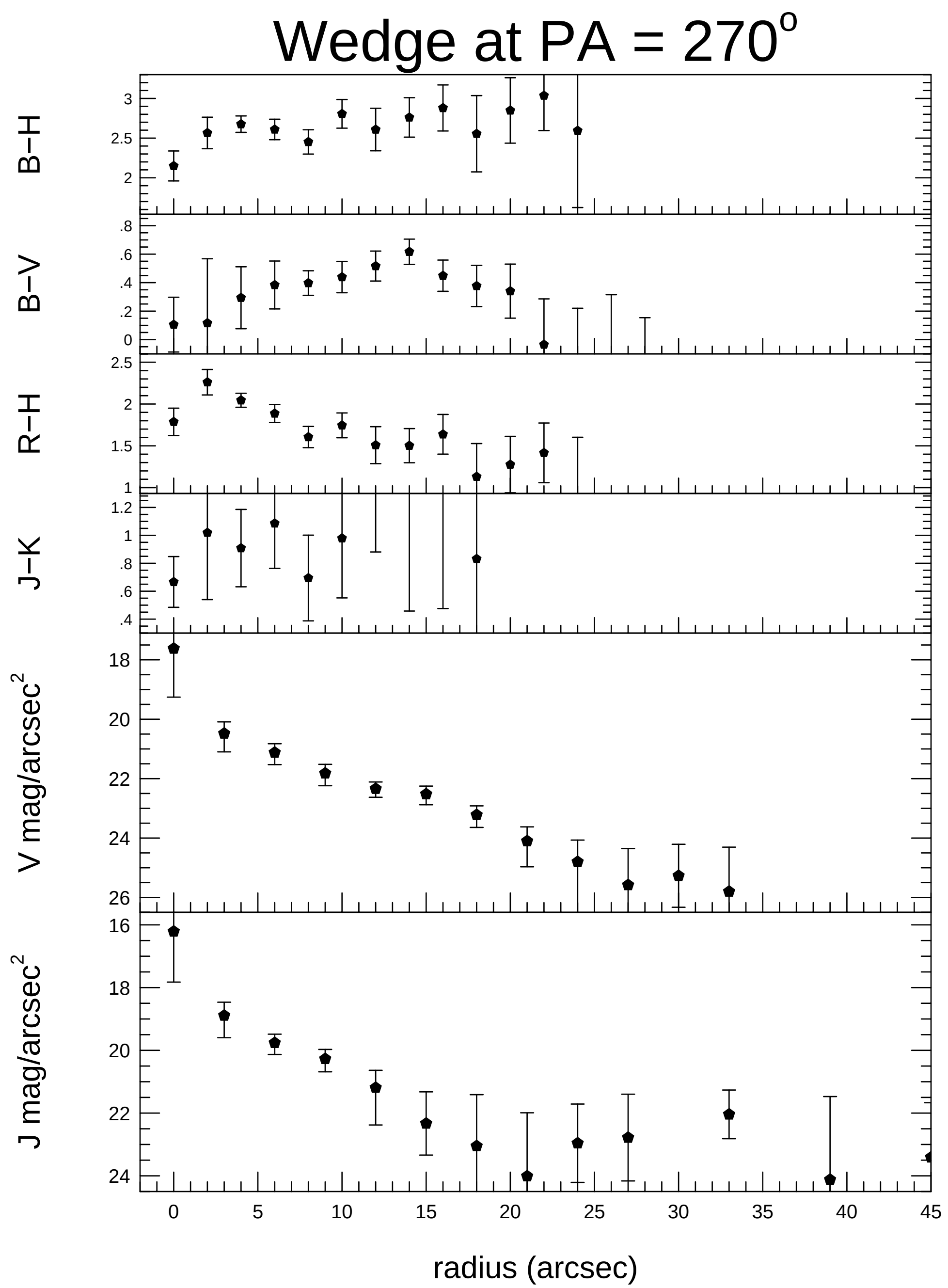
Wedge at PA = 270°
B–H
B–V
R–H
J–K
V mag/arcsec²
J mag/arcsec²
radius (arcsec)

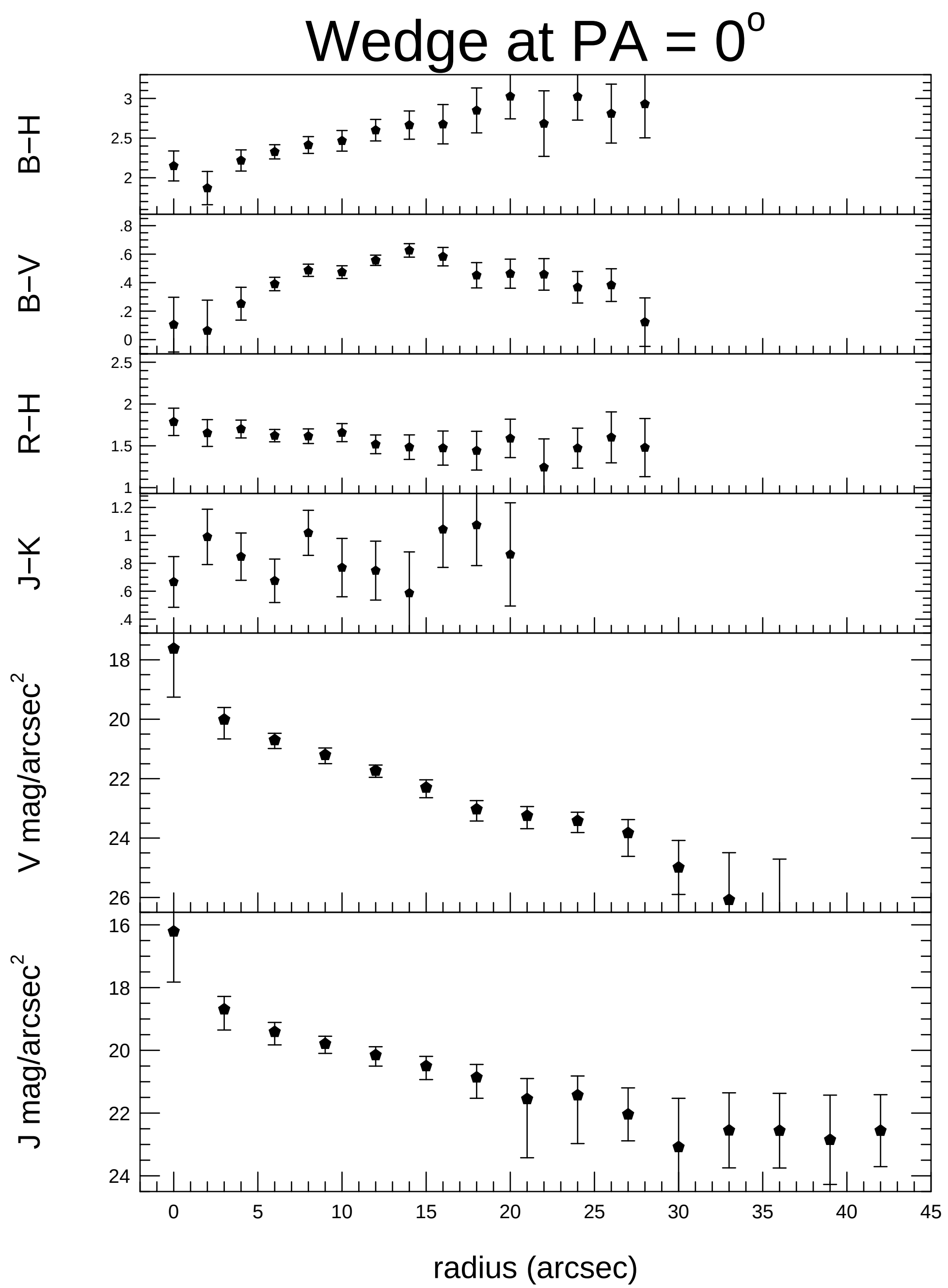

Wedge at PA = 0°
B−H
B−V
R−H
J−K
V mag/arcsec²
J mag/arcsec²
radius (arcsec)
0
5
10
15
20
25
30
35
40
45

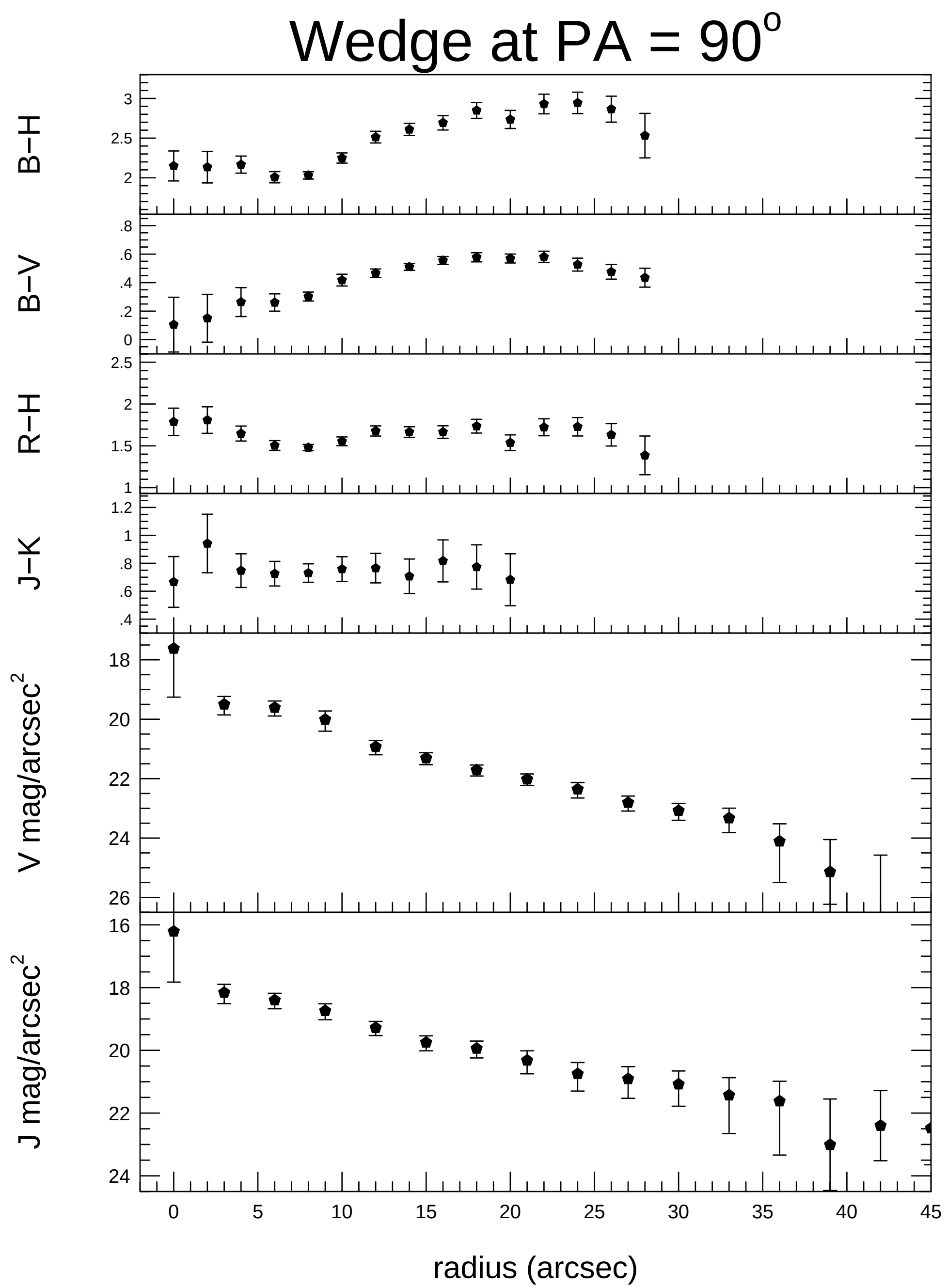
Wedge at PA = 90°
B−H
B−V
R−H
J−K
V mag/arcsec²
J mag/arcsec²
radius (arcsec)

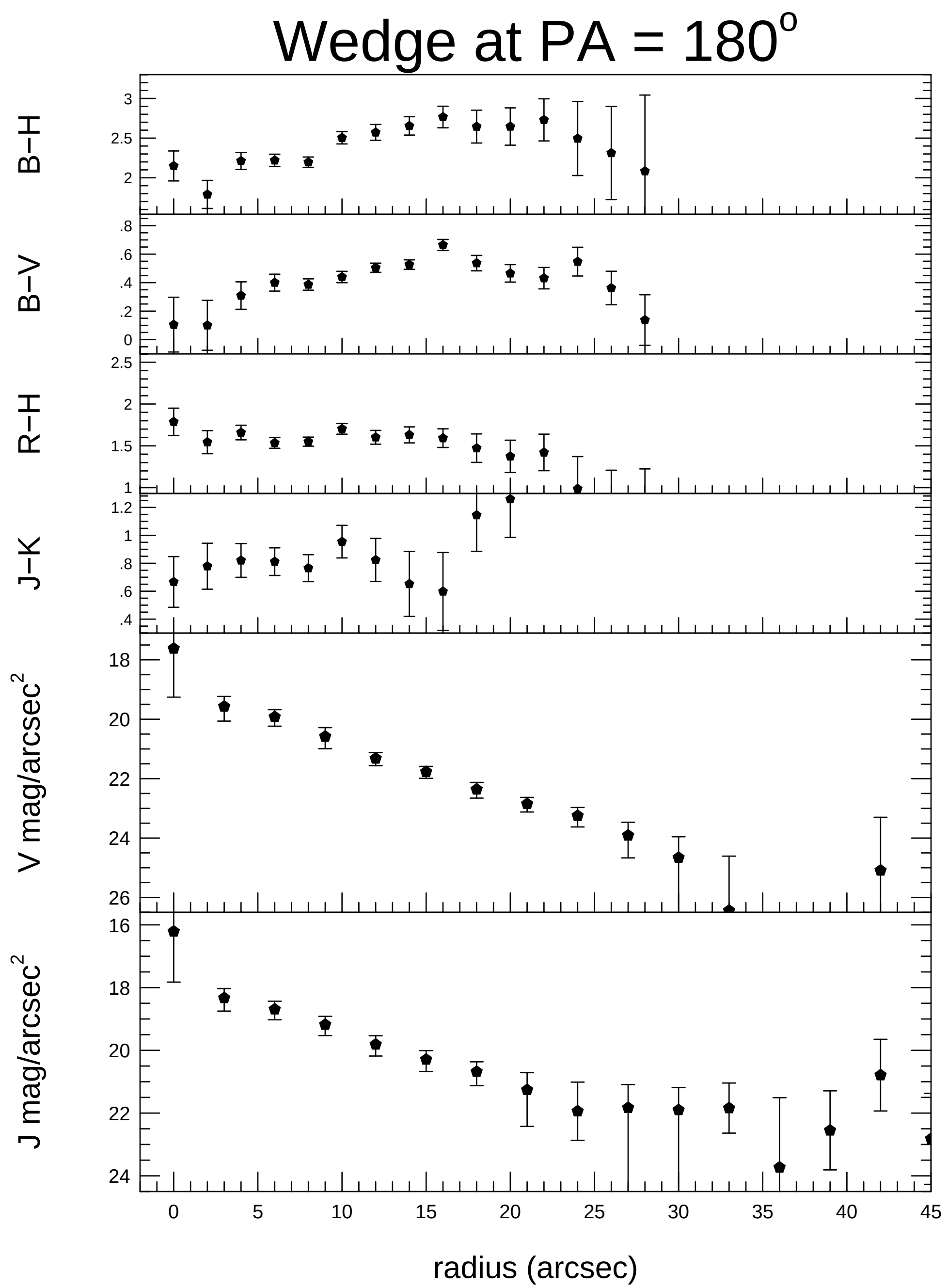

Wedge at PA = 180°
B−H
B−V
R−H
J−K
V mag/arcsec²
J mag/arcsec²
radius (arcsec)
0
5
10
15
20
25
30
35
40
45

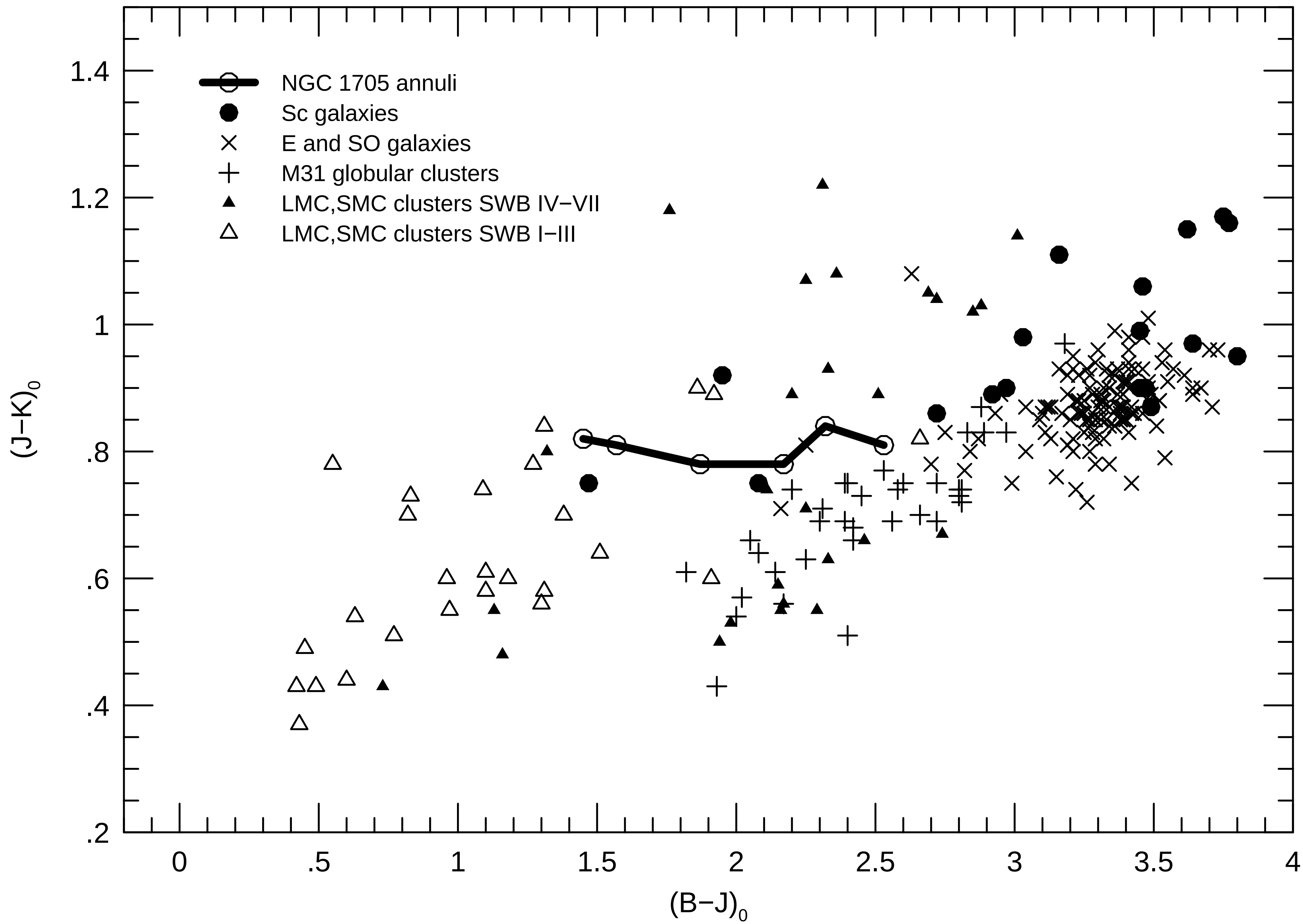
NGC 1705 annuli
Sc galaxies
E and SO galaxies
M31 globular clusters
LMC,SMC clusters SWB IV−VII
LMC,SMC clusters SWB I−III
(J−K)$_0$
(B−J)$_0$
1.4
1.2
1
.8
.6
.4
.2
0
.5
1
1.5
2
2.5
3
3.5
4